\documentclass{article}
\usepackage[utf8]{inputenc}
\usepackage[T1]{fontenc}

\usepackage[a4paper, left = 2cm, right = 2cm, top = 2.5cm, bottom = 3.5cm]{geometry}

\usepackage{graphicx} 
\usepackage{physics}
\usepackage{cmbright}

\usepackage[english]{babel}
\usepackage[parfill]{parskip} 

\usepackage{spreadtab}
\STmessage{false}
\FPmessagesfalse
\usepackage[autolanguage]{numprint}
\usepackage{amsmath}
\usepackage{tabularx}
\usepackage{placeins}
\usepackage{lineno}

\usepackage{authblk}
\usepackage{titlesec}

\titleformat{\section}{\Large}{\thesection}{0.5em}{}
\titleformat{\subsection}{\large}{\thesubsection:}{0.5em}{}
\titleformat{\subsubsection}{}{\thesubsubsection:}{0.5em}{}

\usepackage{cite}

\usepackage{color}
\usepackage[dvipsnames]{xcolor}
\definecolor{KBFIred}{RGB}{163,35,47}
\definecolor{linkblue}{RGB}{7, 140, 255}

\usepackage{url}
\usepackage{hyperref}
\hypersetup{
	colorlinks = true,
	allcolors = linkblue}

\title{\color{KBFIred} \bf Quantum Information meets High-Energy Physics: Input to the update of the European Strategy for Particle Physics}
\author[1]{\bf\normalsize Contact~Persons:~Yoav~Afik \thanks{\href{mailto:yoavafik@gmail.com}{yoavafik@gmail.com}}}
\author[2,3]{\bf\normalsize Federica~Fabbri \thanks{\href{mailto:federica.fabbri@cern.ch}{federica.fabbri@cern.ch}}}
\author[4]{\bf\normalsize Matthew~Low \thanks{\href{mailto:matthew.w.low@gmail.com}{matthew.w.low@gmail.com}}}
\author[5,6]{\bf\normalsize Luca~Marzola \thanks{\href{mailto:luca.marzola@cern.ch}{luca.marzola@cern.ch}}}

\author[7]{~~~~~~~~~~~~~~~~~~~~Juan~Antonio~Aguilar-Saavedra}
\author[8]{Mohammad~Mahdi~Altakach}
\author[9]{Nedaa~Alexandra~Asbah}
\author[10,11]{Yang~Bai}
\author[12]{Hannah~Banks}
\author[13]{Alan~J.~Barr}
\author[7]{Alexander~Bernal}
\author[14]{Thomas~E.~Browder}
\author[15]{Pawe{\l}{}~Caban}
\author[7]{J.~Alberto~Casas}
\author[4]{Kun~Cheng}
\author[16]{Fr\'ed\'eric~D\'eliot}
\author[17]{Regina~Demina}
\author[18,19]{Antonio~Di~Domenico}
\author[20]{Micha{\l}~Eckstein}
\author[21]{Marco~Fabbrichesi}
\author[22]{Benjamin~Fuks}
\author[5,21,23]{Emidio~Gabrielli}
\author[24]{Dorival~Gon\c{c}alves}
\author[25]{Rados{\l}aw~Grabarczyk}
\author[9]{Michele~Grossi}
\author[4]{Tao~Han}
\author[11]{Timothy~J.~Hobbs}
\author[26,27]{Pawe{\l}~Horodecki}
\author[28]{James~Howarth}
\author[29]{Shih-Chieh~Hsu}
\author[30]{Stephen~Jiggins}
\author[30]{Eleanor~Jones}
\author[31]{Andreas~W.~Jung}
\author[32]{Andrea~Helen~Knue}
\author[33]{Steffen~Korn}
\author[34]{Theodota~Lagouri}
\author[2,3]{Priyanka~Lamba}
\author[17]{Gabriel~T.~Landi}
\author[35]{Haifeng~Li}
\author[36]{Qiang~Li}
\author[11,37]{Ian~Low}
\author[2,3,38]{Fabio~Maltoni}
\author[39]{Josh~McFayden}
\author[40]{Navin~McGinnis}
\author[41]{Roberto~A.~Morales}
\author[7]{Jes\'us~M.~Moreno}
\author[42]{Juan~Ram\'on~Mu\~noz~de~Nova}
\author[31]{Giulia~Negro}
\author[3]{Davide~Pagani}
\author[43,44]{Giovanni~Pelliccioli}
\author[45,46]{Michele~Pinamonti}
\author[45,46]{Laura~Pintucci}
\author[9]{Baptiste~Ravina}
\author[36]{Alim~Ruzi}
\author[47]{Kazuki~Sakurai}
\author[48]{Ethan~Simpson}
\author[2,3]{Maximiliano~Sioli}
\author[40]{Shufang~Su}
\author[49]{Sokratis~Trifinopoulos}
\author[14]{Sven~E.~Vahsen}
\author[9]{Sofia~Vallecorsa}
\author[50,51]{Alessandro~Vicini}
\author[52]{Marcel~Vos}
\author[48]{Eleni~Vryonidou}
\author[53]{Chris~D.~White}
\author[54]{Martin~J.~White}
\author[31]{Andrew~J.~Wildridge}
\author[4]{Tong~Arthur~Wu}
\author[55]{Laura~Zani}
\author[29]{Yulei~Zhang}
\author[56]{Knut~Zoch}

\affil[1]{Enrico Fermi Institute, University of Chicago, Chicago, Illinois 60637, USA}

\affil[2]{Dipartimento di Fisica e Astronomia, Universit\`{a} di Bologna, Via Irnerio 46, 40126 Bologna, Italy}

\affil[3]{INFN, Sezione di Bologna, Via Irnerio 46, 40126 Bologna, Italy}

\affil[4]{Pittsburgh Particle Physics, Astrophysics, and Cosmology Center, Department of Physics and Astronomy, University of Pittsburgh, Pittsburgh, USA}

\affil[5]{Laboratory of High-Energy and Computational Physics, KBFI, R\"avala pst 10, 10143 Tallinn, Estonia}

\affil[6]{Institute of Computer Science, University of Tartu, 
Narva mnt 18, 51009 Tartu, Estonia}

\affil[7]{Instituto de F\'\i sica Te\'orica, IFT-UAM/CSIC, c/ Nicol\'as Cabrera 13-15, 28049 Madrid}

\affil[8]{Laboratoire de Physique Subatomique et de Cosmologie (LPSC), Universit\'e
Grenoble-Alpes, CNRS/IN2P3, 53 Avenue des Martyrs, F-38026 Grenoble, France}

\affil[9]{CERN, European Organization for Nuclear Research, Geneva, Switzerland}

\affil[10]{Department of Physics, University of Wisconsin-Madison, Madison, WI 53706, USA}

\affil[11]{High Energy Physics Division, Argonne National Laboratory, Lemont, IL 60439, USA}

\affil[12]{Department of Applied Mathematics and Theoretical Physics, University of Cambridge, Wilberforce Road, Cambridge, CB3 0WA, UK}

\affil[13]{Department of Physics, Keble Road, University of Oxford, OX1 3RH, Merton College, Merton Street, Oxford, OX1 4JD}

\affil[14]{Department of Physics and Astronomy, University of Hawaii, 2505 Correa Road, Honolulu, HI, 96822, USA}

\affil[15]{Department of Theoretical Physics, University of {\L}{\'o}d{\'z}, Pomorska 149/153, PL-90-236 {\L}{\'o}d{\'z}, Poland}

\affil[16]{CEA, Université Paris-Saclay, Institut de Recherche sur les lois Fondamentales de l'Univers (IRFU), F-91191 Gif sur Yvette Cedex, France}

\affil[17]{Department of Physics and Astronomy, University of Rochester, 206 Bausch and Lomb Hall, Rochester, NY}

\affil[18]{Dipartimento di Fisica, Sapienza Universit\`a di Roma}

\affil[19]{INFN Sezione di Roma, P.\ le A.\ Moro 2, I-00185, Rome, Italy}

\affil[20]{Institute of Theoretical Physics, Faculty of Physics, Astronomy and Applied Computer Science, Jagiellonian University, ul. Lojasiewicza 11, 30–348 Kraków, Poland}

\affil[21]{INFN, Sezione di Trieste, Via Valerio 2, I-34127 Trieste, Italy}

\affil[22]{Laboratoire de Physique Th\'eorique et Hautes \'Energies (LPTHE), UMR 7589, Sorbonne Universit\'e et CNRS, 4 place Jussieu, 75252 Paris Cedex 05, France}

\affil[23]{Physics Department, University of Trieste, Strada Costiera 11, I-34151 Trieste, Italy}

\affil[24]{Department of Physics, Oklahoma State University, Stillwater, OK, 74078, USA}

\affil[25]{Department of Physics, University of Oxford, Keble Road, Oxford, OX1 3RH, UK}

\affil[26]{International Centre for Theory of Quantum Technologies, University of Gdańsk, Wita Stwosza 63, 80-308 Gdańsk, Poland}

\affil[27]{Faculty of Applied Physics and Mathematics, National Quantum Information Centre, Gdańsk University of Technology, Gabriela Narutowicza 11/12, 80-233 Gdańsk, Poland}

\affil[28]{SUPA - School of Physics and Astronomy, University of Glasgow, Glasgow, UK}

\affil[29]{Department of Physics, University of Washington, 1410 NE Campus Parkway, Seattle, WA, USA}

\affil[30]{Deutsches Elektronen-Synchrotron DESY, Hamburg, and Zeuthen, Germany}

\affil[31]{Department of Physics and Astronomy, Purdue University, West Lafayette, IN 47906, USA}

\affil[32]{Department of Physics, University of Dortmund,
Otto-Hahn-Str. 4a, Dortmund, Germany}

\affil[33]{I. Physikalisches Institut, Georg-August-Universität Göttingen, Göttingen, Germany}

\affil[34]{Department of Physics, Yale University, New Haven CT, USA}

\affil[35]{Institute of Frontier and Interdisciplinary Science and Key Laboratory of Particle Physics and Particle Irradiation (MOE), Shandong University, Qingdao, China}

\affil[36]{State Key Laboratory of Nuclear Physics and Technology, School of Physics, Peking University, Beijing, 100871, China}

\affil[37]{Department of Physics and Astronomy, Northwestern University, Evanston, IL 60208, USA}

\affil[38]{Centre for Cosmology, Particle Physics and Phenomenology (CP3), Universit\'{e} Catholique de Louvain, B-1348 Louvain-la-Neuve, Belgium}

\affil[39]{Department of Physics and Astronomy, University of Sussex, Brighton, United Kingdom}

\affil[40]{Department of Physics, University of Arizona, Tucson, Arizona 85721, USA}

\affil[41]{IFLP, CONICET - Departamento de F\'isica, Universidad Nacional de La Plata, C.C. 67, 1900 La Plata, Argentina}

\affil[42]{Departamento de F\'isica de Materiales, Universidad Complutense de Madrid, E-28040 Madrid, Spain}

\affil[43]{Dipartimento di Fisica, Università degli Studi di Milano-Bicocca, Piazza della Scienza 3, 20161 Milano, Italy}

\affil[44]{INFN Sezione di Milano-Bicocca, Piazza della Scienza 3, 20161 Milano, Italy}

\affil[45]{Dipartimento Politecnico di Ingegneria e Architettura, University of Udine, Via delle Scienze 206, 33100 Udine, Italy}

\affil[46]{INFN Sezione di Trieste, Gruppo Collegato di Udine, Italy}

\affil[47]{Institute of Theoretical Physics, Faculty of Physics, University of Warsaw, ul.~Pasteura 5, PL-02-093 Warsaw, Poland}

\affil[48]{Department of Physics and Astronomy, University of Manchester, Oxford Road, Manchester M13~9PL, United Kingdom}

\affil[49]{Center for Theoretical Physics, Massachusetts Institute of Technology, Cambridge, MA 02139, USA}

\affil[50]{Dipartimento di Fisica Universit\`a degli Studi di Milano, Via Celoria 16 20133 Milano}

\affil[51]{INFN, Sezione di Milano, Via Celoria 16 20133 Milano}

\affil[52]{IFIC, Universitat de Val\`encia and CSIC, c./ Catedr\'atico Jos\'e Beltr\'an 2, E-46980 Paterna, Spain}

\affil[53]{Centre for Theoretical Physics, School of Physical and Chemical Sciences, Queen Mary University of London, 327 Mile End Road, London E1 4NS, UK}

\affil[54]{ARC Centre for Dark Matter Particle Physics, Department of Physics, University of Adelaide, Adelaide, SA 5005, Australia}

\affil[55]{INFN Sezione di Roma Tre, I-00146 Roma, Italy}

\affil[56]{Laboratory for Particle Physics and Cosmology, Harvard University, Cambridge, 02138 MA, USA}

\date{}

\begin{document}

\maketitle

\bigskip
\bigskip

\bigskip

\newpage

\textbf{Abstract:}
Some of the most astonishing and prominent properties of Quantum Mechanics, such as entanglement and Bell nonlocality, have only been studied extensively in dedicated low-energy laboratory setups. The feasibility of these studies in the high-energy regime explored by particle colliders was only recently shown and has gathered the attention of the scientific community. For the range of particles and fundamental interactions involved, particle colliders provide a novel environment where quantum information theory can be probed, with energies exceeding by about 12 orders of magnitude those employed in dedicated laboratory setups. Furthermore, collider detectors have inherent advantages in performing certain quantum information measurements, and allow for the reconstruction of the state of the system under consideration via quantum state tomography. Here, we elaborate on the potential, challenges, and goals of this innovative and rapidly evolving line of research and discuss its expected impact on both quantum information theory and high-energy physics. 

\bigskip
\bigskip

\tableofcontents

\thispagestyle{empty}
\newpage
\setcounter{page}{1}

\section{Scientific context}
\label{sec:sc}

Entanglement, the hallmark and one of the most perplexing aspects of Quantum Mechanics (QM)~\cite{Einstein:1935rr,Schrodinger1935}, has been observed in a variety of systems ranging from photon pairs to macroscopic objects~\cite{Aspect1982,Hagley1997,Steffen2006,Pfaff2013,Belle:2007ocp,Julsgaard2001,Lee2011,Ockeloen2018,Storz:2023jjx,KLOE:2006iuj,KLOE-2:2021ila}. 
In the presence of entanglement, the full description of a composite system cannot be derived from the individual descriptions of its subsystems because of quantum correlations that interconnect them. 
These correlations may even survive once the subsystems have been spatially separated. 
In the presence of these correlations, it may also happen that the probabilities of obtaining certain outcomes from independent measurements performed on the composing subsystems no longer follow the known factorization rule. 
This phenomenon, known as Bell nonlocality, is quantified by a Bell inequality~\cite{Bell:1964kc}, which is violated in nonlocal theories.
In stark contrast, this inequality is satisfied in any local realistic theory that seeks to remedy the quirkiness of QM through the introduction of hidden variables.
Bell nonlocality is central to Quantum Information Theory (QIT)~\cite{Brunner:2013est}, and its significance was highlighted with the 2022 Nobel Prize in Physics, awarded for experimental tests with entangled photons that demonstrated that Bell inequalities are violated~\cite{aspect1981,Aspect1982,clauser1969,kwiat1995}.

Remarkably, until recently, quantum correlations such as entanglement and Bell nonlocality have not been extensively investigated in the high-energy regime explored in proton collider experiments like the LHC.
The possibility of conducting tests of Bell-type correlations at colliders was first explored in Refs.~\cite{Tornqvist:1980af,Tornqvist:1986pe,Privitera:1991nz,Abel:1992kz,DiDomenico:1995ky,Hiesmayr:2011na}, whereas, more recently, entanglement has been measured in top-antitop-quark pair ($t \bar t$) spin correlations by the ATLAS and the CMS collaborations at the Large Hadron Collider (LHC)~\cite{ATLAS:2023fsd,CMS:2024pts,CMS:2024zkc}. 
Bell nonlocal states have also been observed in data pertaining to spin correlations in charmonium and $B$-meson decays~\cite{Fabbrichesi:2023idl,Gabrielli:2024kbz,Fabbrichesi:2024rec}.
Furthermore, Bell nonlocality has been measured in flavor oscillations of neutral $B$-meson pairs by the Belle collaboration~\cite{Belle:2007ocp}. In addition, the ALICE collaboration recently measured quantum interference effect at the femtometer scale~\cite{ALICE:2024ife}.

Measuring quantum correlations at particle accelerators is challenging but possible~\cite{Afik:2020onf,Fabbrichesi:2021npl,Barr:2021zcp,Severi:2021cnj,Afik:2022kwm,Aguilar-Saavedra:2022uye,Aguilar-Saavedra:2022wam,Ashby-Pickering:2022umy,Barr:2024djo}, even though the detectors were not originally designed for this purpose. 
However, the high energies involved and the fundamental nature of the collider environment provide a compelling setting for novel QIT measurements and, as a result, collider-based measurements serve as a valuable complement to traditional QIT experiments. 

\begin{figure}[th!]
    \centering
    \includegraphics[width=0.53\textwidth]{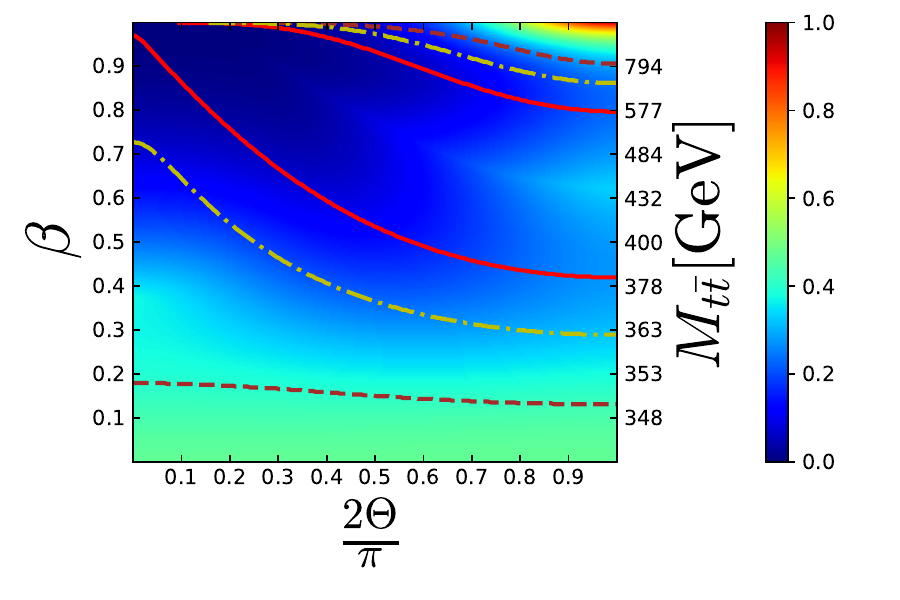}
    \includegraphics[width=0.42\textwidth]{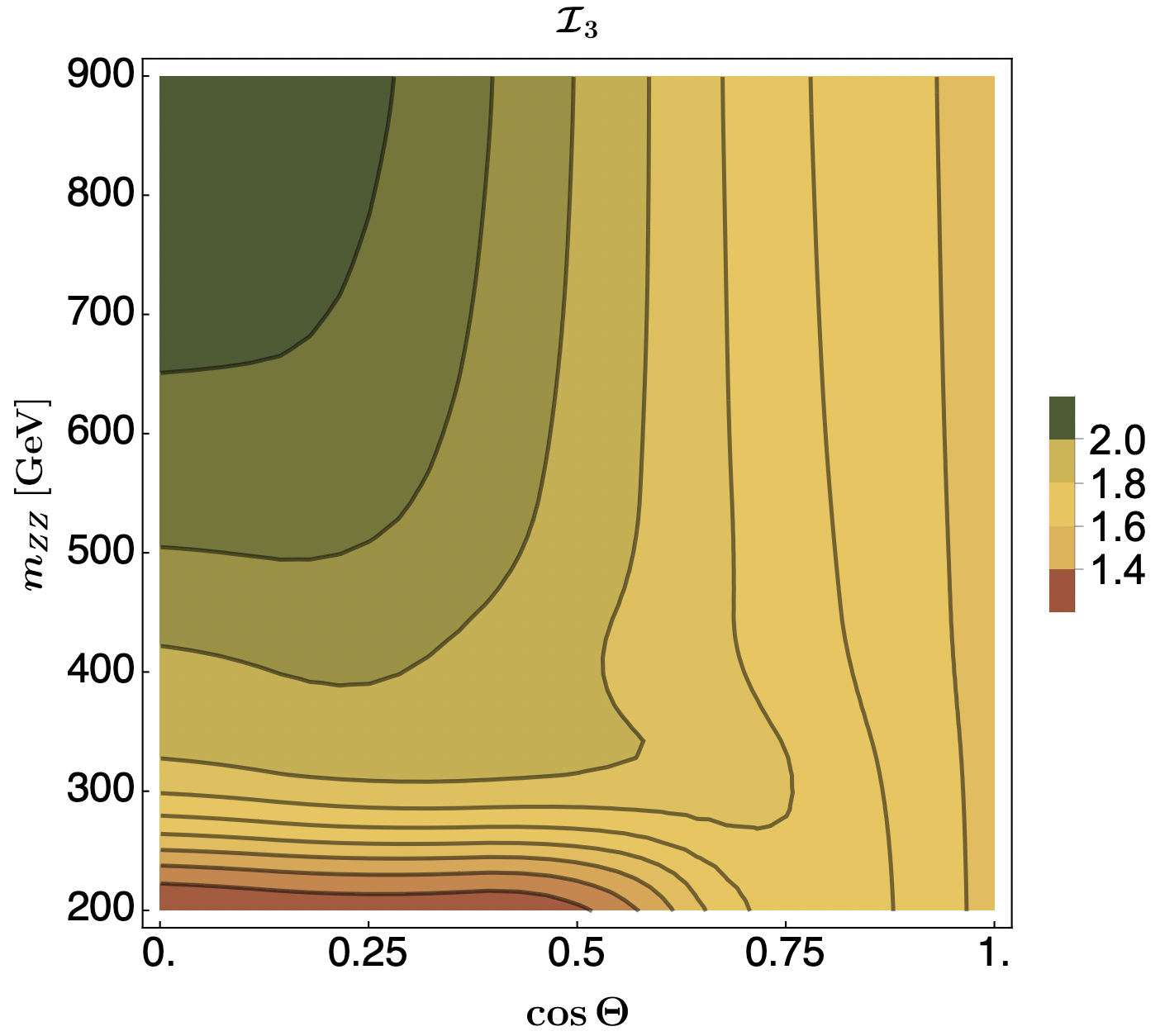}
    \caption{Left panel: Quantum discord of $pp \to t \bar t$ at the LHC as a function of the top velocity $\beta$ and the production angle $\Theta$ in the $t\bar{t}$ center-of-mass frame. Solid red, dashed-dotted yellow, and dashed brown lines are the critical boundaries of separability, steerability, and Bell locality, respectively~\cite{Afik:2022dgh}.
    Right panel: The observable ${\cal I}_3$, where ${\cal I}_3 > 2$ implies a Bell nonlocal state, for the process $pp \to Z Z$ as a function of the invariant mass and scattering angle in the center-of-mass energy frame~\cite{Barr:2024djo}.}
    \label{fig:corr}
\end{figure}

The recent measurements of entanglement in $t \bar t$ by the ATLAS and the CMS collaborations at the LHC have paved the way for this line of studies, serving as a proof of concept for the feasibility of such measurements. 
Not only these are the highest-energy ever measurements of entanglement, they are also the first ones between quarks and the first ones to probe this feature of QM by utilizing unique fundamental particles and interactions, therefore holding great significance.
It is important to note that there is a whole hierarchy of quantum correlations that can be studied. 
For instance, discord is a measure of non-classical correlations that can interconnect the components of a system even if they are not entangled; steering, instead, refers to a property of the state where a measurement performed on one subsystem can steer the quantum state of the other~\cite{Afik:2022dgh}. 
In Fig.~\ref{fig:corr}, the quantum correlations at the LHC in $pp \to t \bar t$ production are shown on the left panel, and Bell nonlocality in $pp \to ZZ$ on the right panel.

The hierarchy of quantum correlations and their possible implications have not yet been studied extensively in the high-energy regime. 
The collider environment carries then a unique opportunity to complete this investigation owing also to the capability of measuring quantum correlations already with present detector technologies. 
For example, measurements such as quantum discord and steering require large data samples. These are already accessible in specific processes, such as top-quark pair final states, within the datasets already collected at the LHC and those expected in upcoming runs. In contrast, achieving comparable statistics in standard experimental setups remains challenging.
Specifically, quantum discord can be measured according to the original definition~\cite{Ollivier2001},
which is extremely challenging in low-energy experiments.
In addition, the steering ellipsoid, which captures an important amount of information about the system~\cite{Jevtic2014}, can be experimentally reconstructed. Previously this was achieved only once by using $5\times 10^4$ detection events~\cite{Zhang2019}.
Furthermore, QM predicts a maximal value for the violation of Bell inequalities, the Cirel'son bound~\cite{Cirelson1980}, which could be probed at the highest available energies. 
Exceeding the bound would demonstrate the existence of a more fundamental theory where nonlocal effects are stronger than in QM~\cite{Brunner:2013est,Popescu:2014wva,Eckstein:2021pgm}. 
Finally, direct access to relativistic and massive particles allows the investigation of quantum systems formed by qubits (such as fermions), and qutrits (such as massive gauge bosons), and may also be valuable for investigating the relativistic properties of the spin operator, a fundamental yet unresolved question in QM and in QIT itself~\cite{Czachor1997,Gingrich2002,Peres2004,Friis2013,Rembielinski2019,Giacomini2019,Taillebois_2021}.

Beyond the fundamental interest in testing quantum correlations at high energies, exploring these concepts could bring several benefits to High-Energy Physics (HEP). Notably, new and promising techniques targeting physics beyond the Standard Model (SM) have already emerged~\cite{Aoude:2022imd,Fabbrichesi:2022ovb,Severi:2022qjy,Maltoni:2024tul,Duch:2024pwm,Sullivan:2024wzl}. 
Additionally, QIT methods can be utilized to explore non-perturbative aspects of Quantum Chromodynamics (QCD)~\cite{Baker:2017wtt,Tu2020,Gong:2021bcp,Florio:2023dke,Barata:2023jgd,Khor:2023xar,Hentschinski:2024gaa,Bloss:2025ywh}.
Furthermore, this research has sparked important discussions and advancements in modeling SM processes, for instance, concerning the production of $t \bar t$ systems used to measure quantum correlations: exploring these concepts has provided new insights into non-relativistic QCD effects, such as toponium formation~\cite{Fuks:2021xje,Aguilar-Saavedra:2024mnm,Fuks:2024yjj,Djouadi:2024lyv}.  Accurate modeling of these effects is also crucial for achieving precision measurements within the SM and, without leveraging variables sensitive to the entanglement between the spins of the top quarks, such effects would remain inaccessible given current reconstruction resolution limits.
The Higgs boson also plays a crucial role in studying quantum correlations, serving as a source of maximally entangled qubits or qutrits that can be analyzed using QIT methods. Moreover, these studies offer additional tools for exploring Higgs couplings and $CP$ structure.~\cite{Bernal:2023ruk,Altakach:2022ywa}

Currently, exploring QIT concepts in collider processes is a highly active area of research, attracting significant attention. 
As a result, a dedicated community has been forming and has been bringing together researchers from both experimental and theoretical backgrounds. Given the rapid development of this innovative research direction and its immense potential in both the near and distant future, this document outlines its anticipated role in shaping the future of HEP.

\section{Objectives} 

Our proposal is to foster the development of a new research area set at the interface between QIT and HEP, with the purpose of further establishing the related multidisciplinary study of fundamental phenomena. From the perspective of QIT, particle accelerators offer a novel environment where the workings of the framework can be probed in a range of energies and in the presence of fundamental interactions that are well beyond the limit of traditional QIT experiments. From the perspective of particle physics, instead, the proposed research promises to extend the reach and improve the sensitivity of current and future searches through QIT techniques, for instance, quantum state tomography, that can effectively be repurposed to study particle phenomena. It is also possible that the adoption of QIT language for the treatment of particle physics problems could encourage experts from the fields of QIT and quantum computing to work on HEP topics, thereby facilitating the transfer of knowledge between these disciplines. 
In particular, this aligns well with the CERN Quantum Technology Initiative~\cite{DiMeglio:2021min} and it could be pivotal for gauging the actual possibilities offered by quantum computers within the field of fundamental research.
    
A meaningful first step for these activities is already offered by the available collider data, which could be scrutinized with QIT methods not only to determine the possibility of measuring phenomena central to this discipline, such as entanglement, but also to assess how QIT methods fare when tasked with highlighting possible deviations from SM predictions. In the following, we detail the concrete objectives that we believe should be pursued in current and future HEP experiments.

\begin{itemize}
\item \textbf{A fundamental study of entanglement.} \\
Within QM, the Hilbert space of multipartite quantum systems is given by the tensor product of the Hilbert spaces describing the composing subsystems. 
In general, such states can involve both classical and quantum correlations.
If the subsystems are entangled, then the multipartite quantum system cannot be written as a simple product of the states describing the composing subsystems.
Entanglement has been studied extensively within QIT using conventional laboratory setups, but not nearly at the same level in relativistic systems.
Furthermore, the properties and the meaning of entanglement between more than two relativistic subsystems are not presently understood.

At collider experiments, the presence of entanglement can be made distinguishable via quantum state tomography, which allows for the full reconstruction of the quantum state of the system under investigation. 
Whereas entanglement is relatively understood and well quantified for bipartite systems comprising components of low dimensionality~\cite{RevModPhys.81.865}, a general characterization is presently missing. 
In relation to that, having access to relativistic multipartite states is crucial to test the relativistic regime of QIT, and it could thus significantly improve our understanding of entanglement by directly probing its nature and properties in this regime. 
Within HEP, this requirement translates into the necessity of final states characterized by a large number of particles generated by a common process, achievable only at a collider able to reach large collision energies~\cite{Sakurai:2023nsc}.

In light of this, FCC-$hh$~\cite{FCC:2018vvp}, a multi-TeV muon collider~\cite{Accettura:2912370}, and the SPPC~\cite{CEPC-SPPCStudyGroup:2015csa} seem the ideal tools to pursue this line of studies. Precision machines such as FCC-$ee$~\cite{FCC:2018evy}, LEP~III~\cite{Blondel:2012ey}, CEPC~\cite{CEPCStudyGroup:2023quu}, or the proposed linear colliders instead~\cite{Behnke:2013xla,Boland:2210892,Nanni:2023yne}, could be used to achieve extremely accurate tomographic measurements that could help to highlight any deviation from the SM predictions. On a more theoretical ground, we also remark that the literature offers several entanglement measures, monotones and witnesses apt to investigate the phenomenon~\cite{Hollands:2017dov}. Their meaning and the utility within particle physics is to be investigated and clarified. 

\item \textbf{HEP for QIT.} \\ 
Particles are inherently quantum systems, so they can be employed as carriers of quantum information.
In fact, any feature modeled with a compact Lie group naturally yields a discrete spectrum that makes particles work as the $d$-level systems, qudits, that are the building blocks of QIT. Information, here, is encoded in properties such as the spin, flavor, or symmetry representation, and this allows us to promptly apply the formalism of QIT to particle physics. The former can then be studied in the setup offered by HEP experiments, which gives access to a range of energies and a variety of interactions well beyond those of traditional QIT experiments which, typically, rely on systems such as laser beams, atomic orbitals and ion traps to probe the workings of the theory~\cite{Bouwmeester1997, Weihs1998, Hensen2015}. HEP then provides a completely new environment into which these investigations can be extended. The study offers a new perspective on the effect that particle interactions, decays, and radiation have on entanglement and decoherence. 
    
Likewise, properties central to QIT that require significant statistics to be determined, such as discord~\cite{Ollivier2001} and the steering ellipsoid~\cite{Wiseman2007}, could be more easily studied at collider experiments. Another quantity of high interest is magic~\cite{Bravyi2005}: a measure of what allows quantum computers to outperform their classical counterparts. Additionally, particle colliders often produce correlated states involving many particles and, therefore, offer the opportunity to test QIT with relativistic multipartite states that are not accessible to low-energy experiments. 
    
The particle physics community has started to ascertain which processes are more suitable for this kind of analysis at present and future colliders; however, the analysis is far from being exhaustive. Presently, the ATLAS and CMS collaborations are forced to use unstable particles to investigate entanglement in spin correlations involving top quarks, bottom quarks, $\tau$-leptons, as well as $W$- and $Z$-bosons. Likewise, similar analyses can be performed with the LHC$b$ detector by reconstructing the decay chains initiated, for instance, by $B$-mesons. Entanglement could also be investigated at LHC experiments through the $B^0 \bar{B}^0$, $D^0 \bar{D}^0$, and $K^0 \bar{K}^0$ flavor oscillations, which offer a complementary probe of quantum behavior~\cite{Takubo:2021sdk}. Stable particles such as electrons, muons and photons could, in principle, also be used for these kinds of analyses provided that the future detector technologies are developed enough to measure their spin directly.

\item \textbf{QIT for HEP.}\\ 
Quantum state tomography at collider experiments is progressively becoming a reality that could deliver entirely new insights in the quantum system under study, through reconstruction of the density matrix of its internal quantum degrees of freedom. 
Besides allowing easy access to observables central to QIT, the density matrix itself is often the best tool to probe for differences between theoretical predictions and the experiment. QIT offers several tools apt to compare density matrices, among them the trace distance~\cite{Helstrom1967} and the fidelity~\cite{Uhlmann1994, Jozsa1994}, which could be repurposed to concisely characterize new physics. Examples that have already been investigated with QIT methods include new effects that modify only spin correlations~\cite{Duch:2024pwm}, modifications of the SM interaction vertices~\cite{Fabbrichesi:2023jep, Bernal:2023ruk, Fabbrichesi:2024xtq, Fabbrichesi:2024wcd, Fabbrichesi:2025ywl}, and additional interactions mediated by a scalar or a pseudoscalar component, used also to probe the  $pp \to t \bar t$ toponium formation~\cite{Fuks:2021xje,Maltoni:2024tul,Fuks:2024yjj}.

On more general grounds, we therefore expect that the application of QIT methods to HEP problems will foster a plethora of new techniques, complementing and possibly extending the reach of more traditional frameworks such as SMEFT and HEFT. 
It is also possible that QIT techniques could help clarify aspects of the SM and quantum field theory. 
For instance, the study of entanglement in spin correlations has already opened a new window to probe off-shell effects in heavy gauge boson decays~\cite{Aguilar-Saavedra:2022wam, Fabbrichesi:2023cev, Bernal:2023ruk}. On top of this, we expect that the adoption of QIT methods for the study of HEP will also encourage the development of more sophisticated tools for the characterization of particle physics processes.

Yet another QIT technique -- quantum process tomography~\cite{Chuang:1996hw} -- can offer unprecedented insight into the quantum dynamics at high energies~\cite{Altomonte:2024upf}. It allows us not only to compare and contrast the predictions of SM with BSM models, but also to test the very foundations of QM, which asserts that quantum dynamics ought to be described by linear completely positive maps~\cite{Eckstein:2021pgm}.

\end{itemize}

\section{Methodology}
\label{sec:methodology}

Currently, quantum observables are measured in colliders as spin correlations between unstable particles. When a particle decays sufficiently quickly, its spin information is transferred to its decay products. In this way, by measuring angular distributions between the decay products in the rest frames of the parent particles, the correlations between the original particles can be deduced, and the full spin density matrix can be reconstructed. 
Such measurements require precise reconstruction of the system to accurately characterize quantum effects. Additionally, it is beneficial to isolate regions of the phase space of the decaying particles to enhance the quantum correlations in the measured density matrix, which often creates more challenges.

Having reconstructed the density matrix, we can study its QIT properties, such as entanglement. Oftentimes, however, quantities can be measured more directly without needing to perform full quantum state tomography first. This is the case for entanglement, which, near production threshold for $t\bar{t}$, is given by the angular difference between the lepton, measured in the antitop rest frame, and the antilepton, measured in the top rest frame.  From this angular distribution, the parameter $D$ can be extracted~\cite{Afik:2020onf} where $D < -1/3$ indicates an entangled state.

\begin{figure}[th!]
    \centering
    \includegraphics[width=0.50\textwidth]{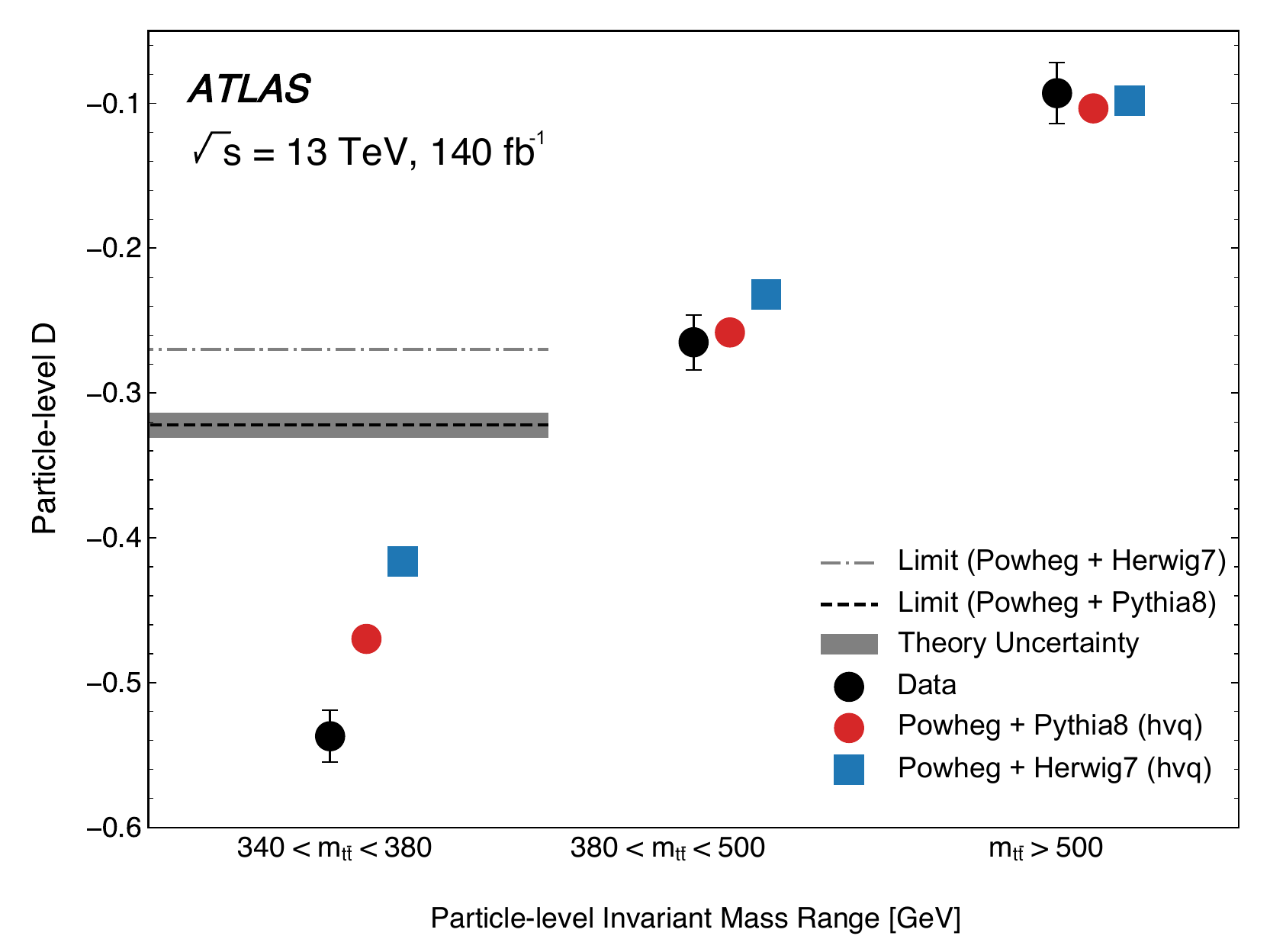}
    \qquad
    \includegraphics[width=0.41\textwidth]{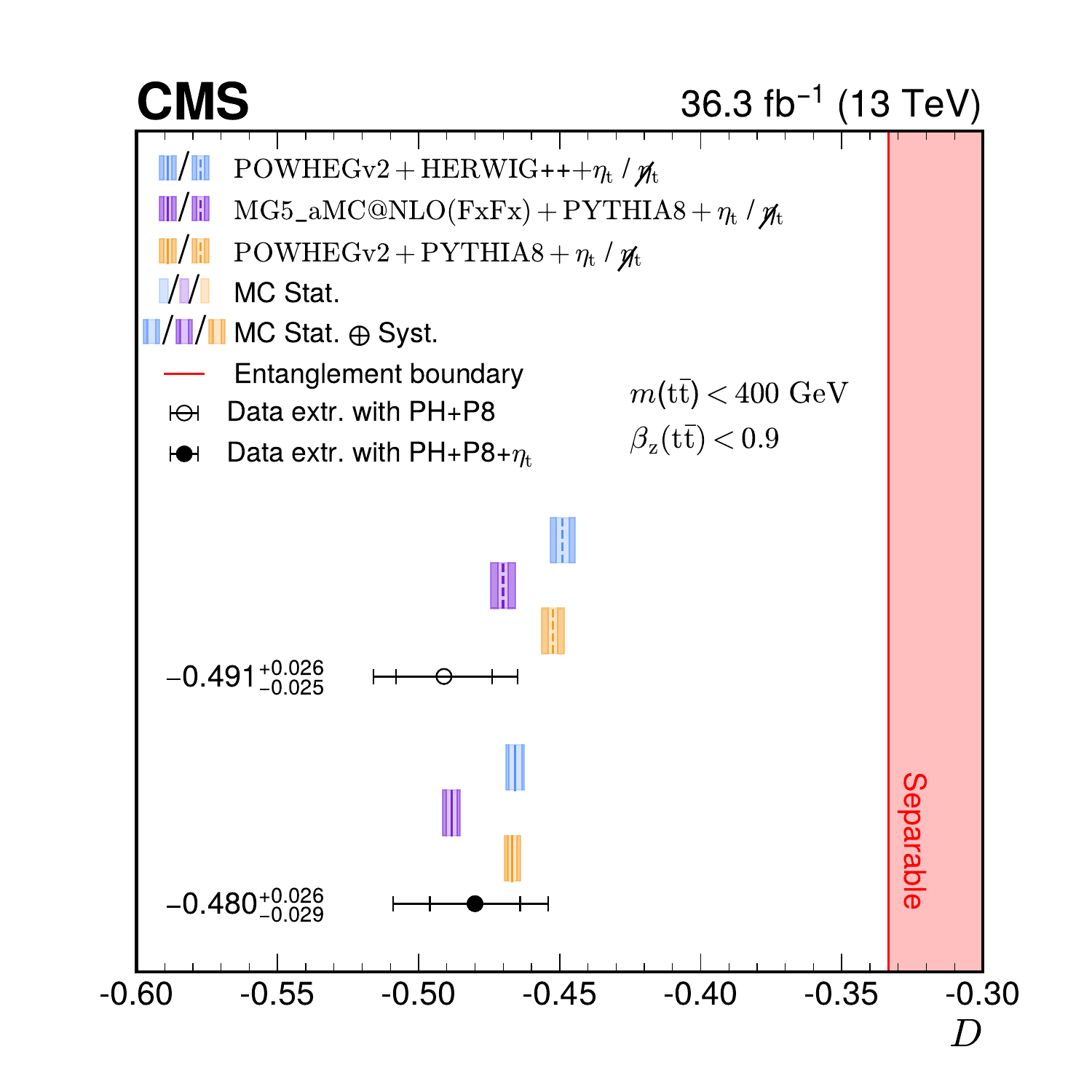}
    \caption{Measurement of the entanglement marker $D$, where $D<-1/3$ indicates entanglement.  Left panel: ATLAS particle-level $D$ measurement compared with various MC models.
    Error bars represent all uncertainties included.
    The entanglement limit shown in the low $m_{t \bar t}$ region is a conversion from its parton-level value of $D = -1/3$ to the corresponding value at particle-level~\cite{ATLAS:2023fsd}.
    Right panel: CMS parton-level $D$ measurement either including (black filled point) or not including (black open point) contribution from toponium, compared to MC predictions with (solid line) or without (dashed line) the inclusion of toponium. Inner error bars represent the statistical uncertainty, while the outer error bars represent the total uncertainty for data~\cite{CMS:2024pts}.   
    }
    \label{fig:tt_LHC}
\end{figure}

Both the ATLAS and CMS collaborations have proven they are up to the task of detailed spin correlation measurements and have measured entanglement in the top-pair final state near the production threshold~\cite{ATLAS:2023fsd,CMS:2024pts}, shown in Fig.~\ref{fig:tt_LHC}. 
The CMS collaboration has also measured entanglement in the lepton and jets final state of the top-pair at high $m_{t \bar t}$, and also the polarizations and spin correlations of the $t \bar t$ system~\cite{CMS:2024zkc}.
This result is the tip of the iceberg in terms of QIT observables that can be measured in the near future.  Using spin correlations to perform quantum tomography, we can expect experimental tests of quantum discord~\cite{Afik:2022dgh,Han:2024ugl} in top-pairs and of entanglement and Bell nonlocality in $\tau^+ \tau^-$ events at the LHC~\cite{Fabbrichesi:2022ovb,Zhang:2025mmm}.

The potential physics applications at a future collider are also myriad.
A lepton collider offers especially useful complementarity from its large statistical sample at a fixed collision energy and its nearly vanishing backgrounds.  The full knowledge of the collision kinematics also makes final states with multiple invisible particles, like decays of $\tau^+ \tau^-$, immediately accessible~\cite{Maltoni:2024csn,Fabbrichesi:2024wcd}. Future hadron colliders, with their higher collision energy, would produce multi-particle final states at higher rates, which would enable the study of more complex quantum systems.

There is another method, other than using spin correlations, to reconstruct the density matrix of particle spins.
Each combination of final state particle spins leads to a distinct differential cross section.  Detailed measurements of the kinematics of particles, therefore, reveal the polarizations and spin correlations and consequently the density matrix~\cite{Cheng:2024rxi}.

Flavor oscillations allow for the quantum number of flavor, rather than spin, to form the building blocks of quantum states produced at colliders.  
These oscillations occur between a particle and its antiparticle.  
When a decay occurs, the decay products will identify whether it was the particle or the antiparticle at the time of decay. 
The reconstruction of the decay products can identify whether it was the particle or the antiparticle which decayed. 
Measuring this as a function of the time of the decays allows for the reconstruction of parts of the density matrix. 
In particular, Belle~II~\cite{Belle-II:2018jsg} provides a high-rate production of entangled $B$-meson pairs, offering unique opportunities for entanglement studies that remain largely unexplored. 
Current analyses assume perfect flavor entanglement of $B$-meson pairs, with potential decoherence effects from environmental interactions or new physics mostly untested~\cite{Bertlmann:2001iw}. 
A previous Belle study provided a quantitative test of Bell nonlocality in flavor physics~\cite{Belle:2007ocp}. 
Similar measurements are possible also in the $D^0 \bar{D}^0$ and $K^0 \bar{K}^0$ systems~\cite{Chen:2024drt,KLOE-2:2021ila,Hiesmayr:2011na} and novel effects are under study~\cite{DiDomenico:2023vkf}.

The current methodology has incredible potential to reveal aspects of QIT at high energies.  
Advancements in accelerator and detector technology may further enhance and catalyze the QIT capabilities of HEP experiments. 
Even more exciting possibilities open up if the beams at a future collider could be polarized along any direction. 
This would enable a preparation of the spin state of the colliding particles, shifting the paradigm of collider physics from observations towards experiments with tunable settings. 
When combined with quantum state tomography, polarized beams would enable quantum process tomography~\cite{Altomonte:2024upf}, providing an unprecedented opportunity to explore quantum dynamics at high-energies~\cite{Altomonte:2023mug,Barr:2024djo} and test the linearity of the quantum theory itself~\cite{Eckstein:2021pgm}.

Another technology that would greatly benefit the QIT program at future colliders would be the development of a technique to measure the spins of stable particles directly. 
This would allow spins to be measured along a chosen spin axis defined by the detector setup, and bring the HEP experiments more into line with traditional low-energy experiments, where the spin quantization of stable particles is directly measured.  
One emerging technology that may address this aspect is nitrogen-vacancy centers in diamond~\cite{10.3389/fphy.2022.887738}.

\section{Readiness and expected challenges}

The initial measurements conducted by the ATLAS and CMS collaborations of entanglement and the spin density matrix in final states with top-quark pairs have highlighted key limitations that must be addressed in the coming years. 
Overcoming these challenges is essential for improving the precision of such measurements and expanding their applications, both in terms of measurable quantities and accessible final states.

A fundamental aspect of the quantum state tomography approach described in the previous section involves transforming to the center-of-mass frame and subsequently to the rest frame of the particle of interest. This process necessitates the full reconstruction of the final state kinematics, which has two major implications.  First, all aspects of object reconstruction, including energy calibration, influence the measurement of angular variables that serve as inputs to quantum observables. Second, the presence of neutrinos in the final state, which remain undetectable in multipurpose experiments, introduces a significant challenge. Their presence degrades the resolution not only of the extracted quantum information observables but also of the kinematic variables used to define the phase-space regions where maximal quantum correlations are expected.

In certain cases, the presence of neutrinos precludes or significantly complicates these studies to specific final states, such as $t\bar{t}\to b\bar{b} \ell^{+} \nu \ell^{-} \bar{\nu}$, $H \to \tau^+ \tau^- \to \pi^+ \bar{\nu} \pi^- \nu$, $Z\to \tau^+ \tau^- \to \pi^+ \bar{\nu} \pi^- \nu$, or $H \to WW^{*} \to \ell^+\nu\ell^-\bar{\nu}$. 
An improvement in the reconstruction of the final states including neutrinos can be achieved with advanced  machine learning (ML) techniques~\cite{Leigh:2022lpn}. In addition, the well-defined initial state and the cleaner final state characteristic of lepton colliders, will mitigate these reconstruction challenges and enable access to final states involving multiple neutrinos~\cite{Altakach:2022ywa}.

A different direction to improve the resolution of the final state reconstruction is to use hadronic final states, with a limited number of neutrinos~\cite{Fabbri:2023ncz}. In these final states, however, additional limitations arise.
The correlation between the spin of the parent particle and the direction of its decay products depends on the flavor of the latter. While identifying final-state flavors is straightforward for charged leptons, hadronic final states pose a significant challenge. Currently, no technology exists to efficiently tag the origin of a light jet -- whether it originates from a gluon or an up, down, or strange quark -- and the existing techniques to identify jets originating from charm quarks have limited  efficiencies~\cite{ATLAS:2022qxm}. These limitations affect both the feasibility and the precision of measurements involving jets in the context of quantum information observables. 
In this area, promising developments based on ML are emerging, which could enhance jet flavor identification~\cite{Dong:2024xsg} on one hand and, on the other hand, improve final state reconstruction by accurately matching reconstructed objects with their initial partons~\cite{Shmakov:2021qdz}.

\begin{figure}[th!]
    \centering
    \includegraphics[width=0.40\textwidth]{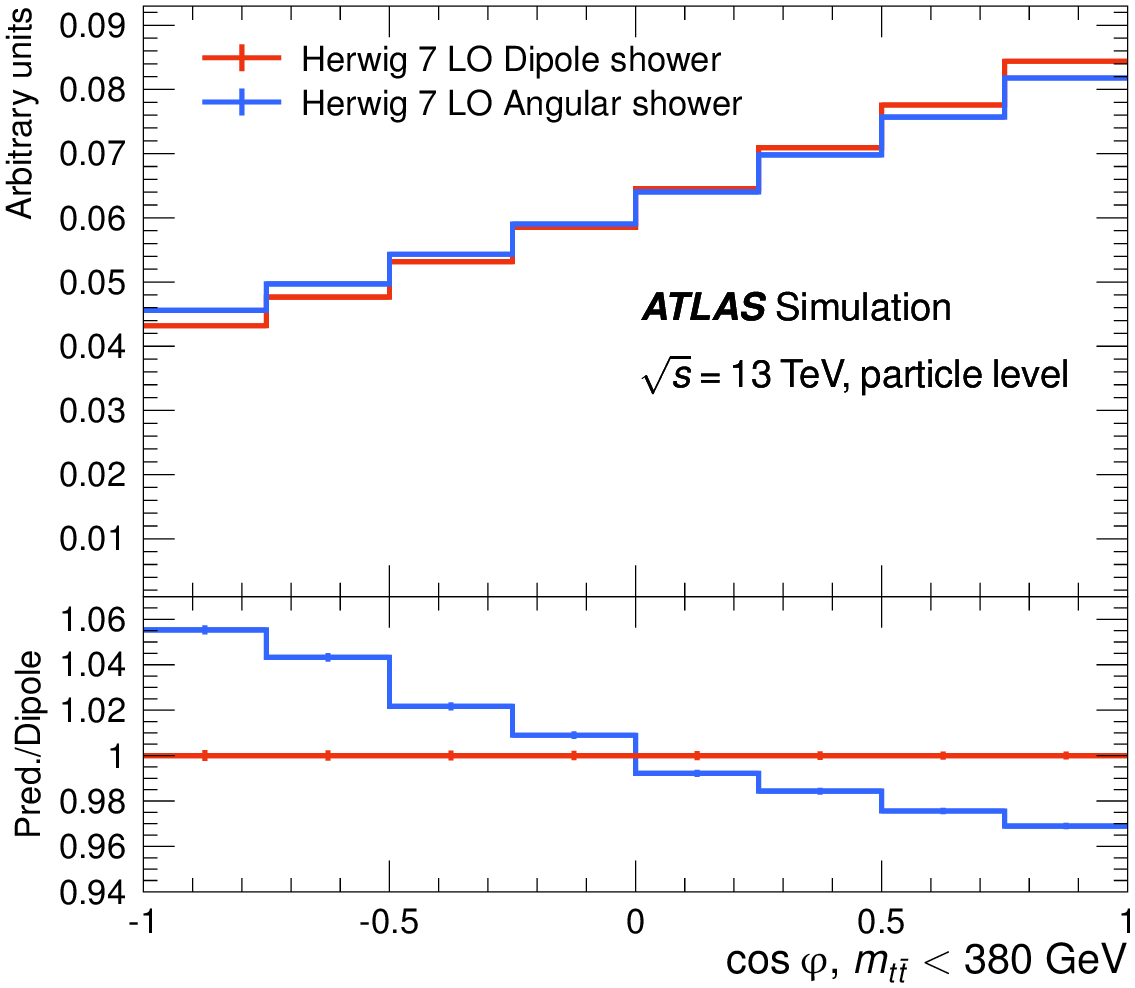}
    \qquad
    \includegraphics[width=0.49\textwidth]{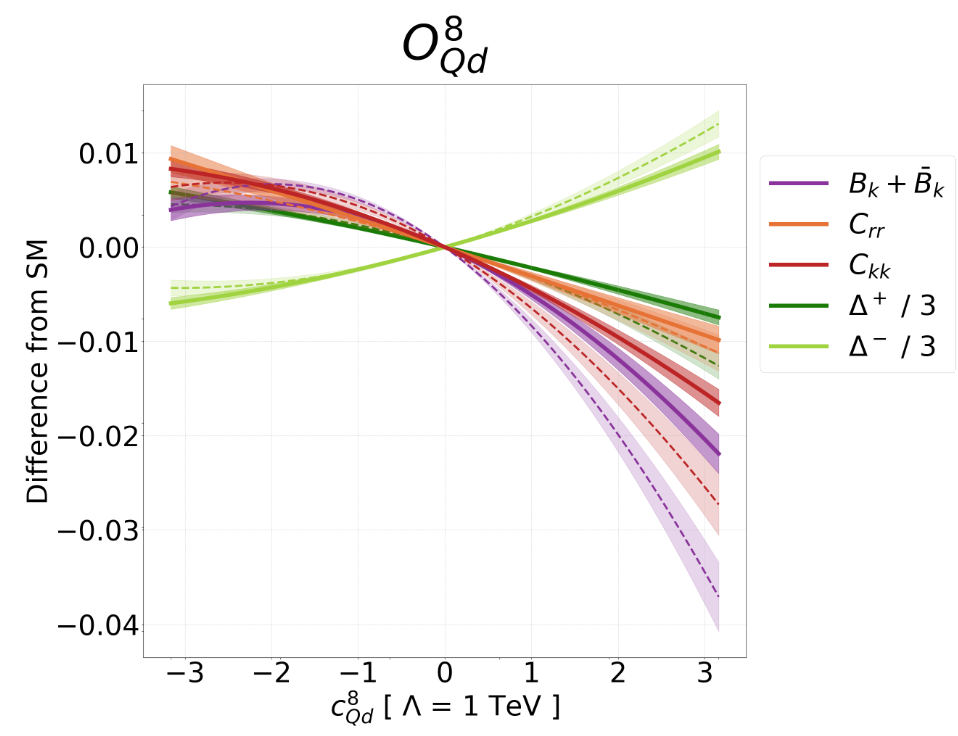}
    \qquad\qquad
    \caption{Left panel: Comparison between two different approaches in the showering algorithm to the simulation of top-quark pair production as a function of the angular variable input to the entanglement witness $D$ calculation~\cite{ATLAS:2023fsd}.
    Right Panel: Difference with respect to the SM prediction of several terms of the spin density matrix and entanglement witnesses ($\Delta^{+},\Delta^{-}$) in top-quark pair production as a function of a coupling used to parametrize the presence of new physics in the SM vertices. The continuous line is obtained using only leading-order simulations, while the dashed line includes higher-order effects in QCD~\cite{Severi:2022qjy}.}
    \label{fig:MCimpact}
\end{figure}

Another limiting factor identified in these initial measurements is the accuracy of Monte Carlo (MC) simulations, and with this, the need to carry out an extended theoretical program to control QCD and related uncertainties. 
The analysis strategies employed rely on the ability of MC simulations to accurately predict detector-level distributions of angular quantities, which serve as inputs for the extraction of the spin density matrix. 
Given the sensitivity of these measurements to all aspects of reconstructed objects, it is imperative that every step of the simulation pipeline is highly precise. 
This includes not only the matrix element calculations but also parton-shower modeling and hadronization. An example of how these aspects affect the simulation in top-quark pair final states of the angular distribution used to extract the quantum entanglement marker $D$ is shown in Fig.~\ref{fig:MCimpact} (left panel). 

With respect to matrix element calculations, early investigations into the impact of higher-order corrections on the extraction of the spin density matrix and related observables have demonstrated their relevance, both in the extracted quantities and their potential to constrain new physics~\cite{Grossi:2024jae,DelGratta:2025qyp,Severi:2022qjy}, as shown in Fig~\ref{fig:MCimpact} (right panel). 
However, such corrections have not yet been investigated in most phenomenological analyses, although several MC event generators already support at least next-to-leading-order QCD corrections to processes of interests. Further development is required to include higher-order electroweak corrections in the MC simulations.
Another related challenge that must be addressed in the coming years is the change in the intrinsic structure of spin density matrices in the presence of higher-order corrections.
Additionally, the simulation of parton-shower effects and the corresponding matching to fixed-order predictions are undergoing continuous refinement. Assessing the impact of parton-shower modeling on quantum-information observables and contributing to its development will be a crucial aspect of future research, with a special focus on the inclusion of exact spin correlations.

Another current issue concerns the possible study of multipartite systems~\cite{Horodecki:2024bgc}, made accessible by multi-particle final states at future colliders characterized by a large value of the center of mass energy. Besides overcoming the limitations imposed by the lower statistics that typically affects the final states of interest, the theoretical understanding of entanglement among three or more parties becomes much more involved than in the simple case of a bipartite system. In order to fully capitalize on the opportunities offered by future colliders, it is therefore mandatory to further develop the fundamental understanding of relativistic multipartite systems and of the related entanglement measures and witnesses. These activities are as crucial as the phenomenological studies aimed to isolate the best final states for such analyses.

A further limitation of existing measurements, which could be mitigated in the near future or at next-generation colliders, is the lack of statistical precision. This issue is particularly pronounced for specific observables such as the steering ellipsoid~\cite{Afik:2022dgh}, for final states involving rare processes, for extensions to multi-particle final states, and for measurements probing extreme regions of phase space at the energy frontier. The latter, for instance, is particularly relevant to testing Bell nonlocality in final states with top-quark pairs.

As outlined in Sect.~\ref{sec:sc}, the application of quantum information techniques to the study of quantum observables is evolving rapidly. This progress includes both theoretical advancements and experimental improvements, where continuous refinements are expected in the coming years.

\section{Timeline}

Several studies have already highlighted the results that can be reached in the next few years.
We elaborate below on important milestones expected to be accomplished, focusing on colliders which are currently the mainstream plan in Europe.
These results are all based on existing technology and methodologies, while new and more stringent results can be achieved with further developments.

\begin{figure}[th!]
    \centering
    \includegraphics[width=0.49\textwidth]{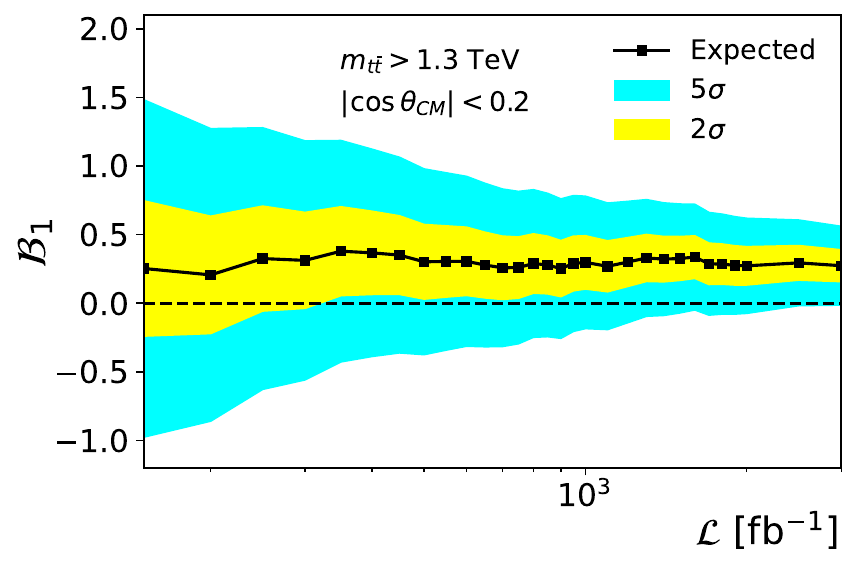}
    \qquad
    \includegraphics[width=0.40\textwidth]{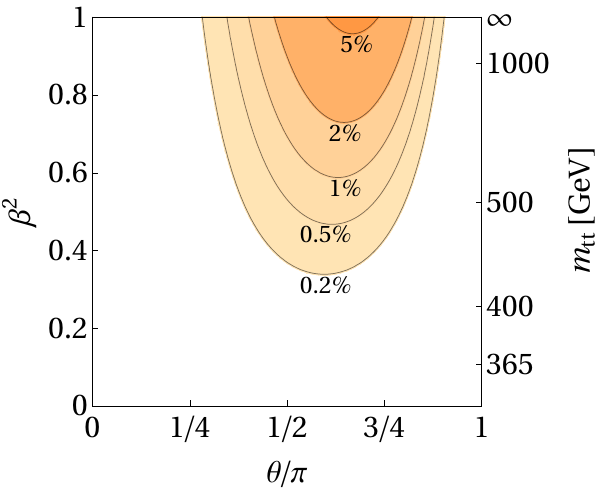}
    \caption{Left panel: Bell indicator $\mathcal{B}_{1}$ as a function of the luminosity in $pp \to t \bar t$ collisions at the LHC and the HL-LHC, where $\mathcal{B}_{1}>0$ implies that the state is Bell nonlocal. The yellow and blue areas represent the regions for an expected statistical significance of $2\sigma$ and $5\sigma$, respectively~\cite{Dong:2023xiw}. 
    Right panel: minimum experimental accuracy estimated to measure Bell nonlocality at $5 \sigma$ in $e^+e^- \to t \bar t$ collisions, as a function of the top velocity squared $\beta^2$ and the production angle $\theta$ in the $t\bar{t}$ center-of-mass frame~\cite{Maltoni:2024csn}.}
    \label{fig:Bells_tt}
\end{figure}

\begin{figure}[th!]
    \centering
    
    \includegraphics[width=0.40\textwidth]{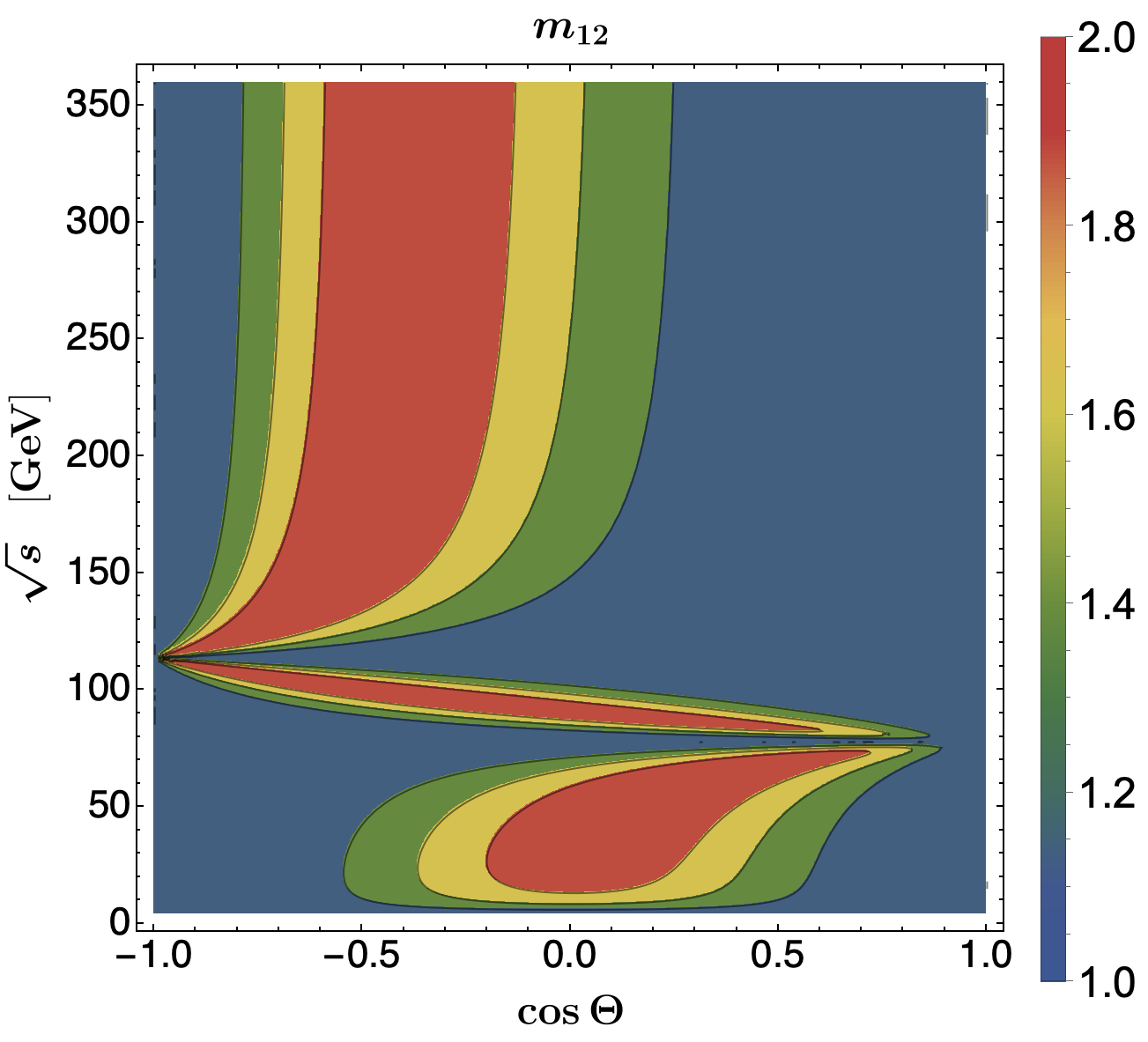}
    \qquad
    \includegraphics[width=0.55\textwidth]{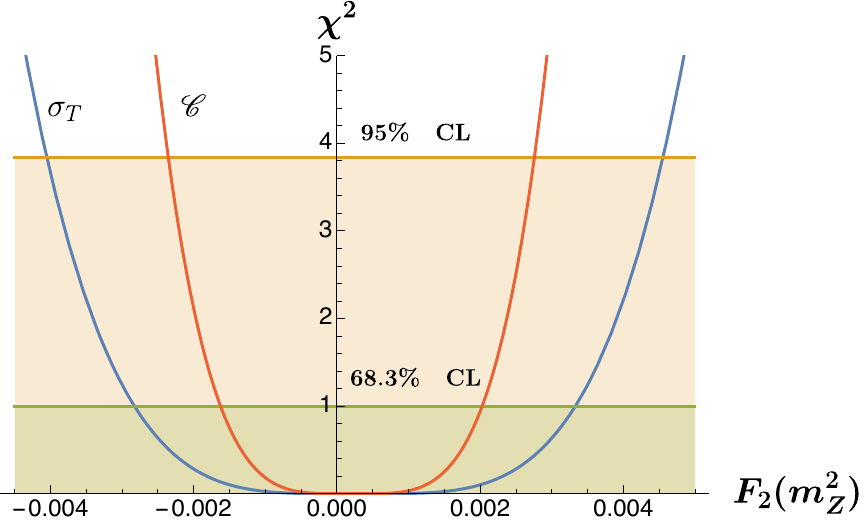}
\caption{
    Left panel: Analytic solution for Bell nonlocality in the parton-level process $e^+ e^- \to \tau^+ \tau^-$. The presence of Bell nonlocality is signaled by $m_{12}>1$.
    Right panel: Tests of $\chi^2$ in $e^+e^-$ events for the form factor $F_2^V(m_Z^2)$, acting as anomalous coupling of the $\tau$-leptons to the $Z$-boson, using the cross section (blue) and entanglement marker (red), show that the latter provides more stringent limits.
    Both plots are from Ref.~\cite{Fabbrichesi:2024wcd}.}
    \label{fig:NP1}
\end{figure}

\begin{itemize}
\item \textbf{LHC.}\\
After the initial observations of entanglement by ATLAS and CMS, further experimental publications are expected, exploiting the already large amount of data collected by the LHC experiments.
Most of these publications follow current suggestions. 
A few of the notable expected measurements to come include studying further quantum correlations in $pp \to t \bar t$, and in new systems like $pp \to H \to VV^*, V = Z,W^\pm$ and $pp \to VV$, among others.
    
In $t \bar t$, it is expected that further observations of entanglement, and quantities like discord, steering, and magic, will be measured in the near future~\cite{Afik:2022dgh,Han:2024ugl,White:2024nuc}. 
Furthermore, a measurement of the post-decay entanglement is possible between one of the top quarks and the $W$-boson from the other top decay~\cite{Aguilar-Saavedra:2023hss}.
This is unique since it allows one to study the evolution of entanglement after the decay of one of the particles, and also allows an opportunity to measure entanglement between a fermion and a massive boson.

Quantum correlations, and in particular Bell nonlocality, have also been explored in Higgs boson decays $pp \to H \to VV^*$ and $pp \to VV$, but current estimations show that while an observation is not yet possible with current data, evidence is expected~\cite{Barr:2021zcp,Aguilar-Saavedra:2022wam}.
Finally, it was shown that entanglement can be also measured in hadronizing systems, in particular in $b \bar b$ production~\cite{Afik:2024uif}. This can be of great interest for the characterization of the quark-gluon plasma, which presents a highly non-trivial spin structure~\cite{STAR2017,ALICE:2021pzu,HADES:2022enx,STAR:2022fan,Giacalone:2025bgm}.

\item \textbf{HL-LHC and Belle II.}\\
The large statistics expected by the HL-LHC are extremely important to be able to observe rare phenomena.
This is true in general, but even more to measure quantum correlations, which are typically maximal in extreme parts of phase space, or for processes with a low cross section.
In particular, observing Bell nonlocal states in $t \bar t$ is extremely challenging~\cite{Dong:2023xiw,Han:2023fci,Cheng:2023qmz,Cheng:2024btk}, since in $t \bar t$ it is only present after imposing extreme cuts on the invariant mass of the top-quark pairs. 
In the left panel of Fig.~\ref{fig:Bells_tt} an example for the expected sensitivity of measuring Bell nonlocality in $t \bar t$ events is shown, reaching almost $5 \sigma$ with the HL-LHC expected data.
    
In Higgs boson decays into two bosons, $H \to VV^*$, it is crucial to have more data available, given the small cross section.
Using current projections, it is likely to be able to establish evidence of Bell nonlocality in $t \bar t$ by the end of the HL-LHC, and an observation in $H \to VV^*$, as the expected statistics should allow it~\cite{Fabbrichesi:2023cev}.

Within the timescale of HL-LHC, additional measurements are expected by Belle~II, a next-generation flavor physics experiment that already began collecting electron-positron collision data~\cite{Belle-II:2018jsg}, with strong European participation~\cite{Belle-II:2025wpi}.
The large dataset expected to be collected will contain billions of decays of bottom mesons, charm hadrons, and $\tau$-leptons.
Belle~II's improved capabilities, such as independent measurement of $B$-meson production times due to the nano-beam scheme, enable more precise tests of flavor entanglement decoherence. 
Additionally, Belle~II can study spin correlations in $\tau$-lepton pairs and could confirm Bell nonlocality~\cite{Ehataht:2023zzt}.

\item \textbf{Future lepton colliders}\\
Currently there are several projects discussing the construction of a future lepton collider. Some prominent examples are: FCC-$ee$~\cite{FCC:2018evy}, LEP~III~\cite{Blondel:2012ey},  CEPC~\cite{CEPCStudyGroup:2023quu}, Muon Colliders~\cite{Accettura:2912370}, ILC~\cite{Behnke:2013xla}, CLIC~\cite{Boland:2210892} and C$^{3}$~\cite{Nanni:2023yne}.
The distinctive processes, the precise knowledge of the center-of-mass energy of the process, and the large statistics expected at lepton colliders offer many opportunities to the field.
In particular, the background for processes like $e^+e^- \to Z/\gamma^* \to \tau^+\tau^-$~\cite{Fabbrichesi:2024wcd}, $e^+e^- \to Z H (\to \tau^+\tau^-)$~\cite{Altakach:2022ywa}, $e^+e^- \to Z H (\to ZZ^*)$~\cite{Wu:2024ovc}, $\mu^+\mu^-\to ZZ$~\cite{Ding:2025mzj} and $\mu^+\mu^-\to \nu_{\mu}\bar{\nu}_{\mu} H (\to ZZ^*)$~\cite{Ruzi:2024cbt} is expected to be small and the resolution to reconstruct the whole final state is expected to be very accurate, allowing precise measurements of quantum correlations such as entanglement, steerability and Bell nonlocality. 
In the right panel of Fig.~\ref{fig:Bells_tt} an example for the required experimental precision to measure Bell nonlocality in $e^+e^- \to t \bar t$ events with $5 \sigma$, as a function of the center-of-mass energy, is shown.
In the left panel of Fig.~\ref{fig:NP1} the regions of phase space in which the quantum state of the system is Bell nonlocal in $e^+e^- \to \tau^+\tau^-$ events, as a function of the center-of-mass energy, are shown.
Another remarkable example is the possibility of measuring, in a spin-entangled pair of particles, that the entanglement can increase after one particle decays~\cite{Aguilar-Saavedra:2024fig}. This phenomenon has no analogue for stable particles, and can be measured with $t \bar t$ at lepton colliders with polarized beams.
The increased precision that will be reached on QIT related measurements at lepton colliders implies also an enlarged sensitivity to investigate new physics effects~\cite{Aoude:2023hxv}, an example is shown in the right panel of Fig.~\ref{fig:NP1}.

Finally, we note that with current detector setups, we can reconstruct the quantum state of the measured system.
Nevertheless, in order to perform genuine Bell tests, measuring the spin of individual particles provides invaluable assistance.
It is likely to expect that detectors which can accomplish this will be developed in the timescale of future lepton colliders. 
This development can also contribute to a deeper investigation of the correct definition of a well-behaved relativistic spin operator, a fundamental yet unresolved question in QM~\cite{Czachor1997,Gingrich2002,Peres2004,Friis2013,Rembielinski2019,Giacomini2019}.
Furthermore, the option to polarize the beams in lepton colliders would open up the possibility of new foundational tests of QM~\cite{Eckstein:2021pgm,Altomonte:2024upf}.

\item \textbf{Future hadron colliders.}\\
So far, most of the theoretical studies focused on bipartite systems.
One of the main reasons for this is that at the LHC, and at future lepton colliders, the cross section for the production of more than two heavy particles that decay quickly enough to propagate their spin information to their decays products, is small.
Nevertheless, in new machines targeting a center-of-mass energy of the order of a hundred TeV, such as FCC-$hh$~\cite{FCC:2018vvp} or SPPC~\cite{CEPC-SPPCStudyGroup:2015csa}, these processes will be more abundantly produced.
This presents an opportunity to study entanglement in multi-particle systems, a challenging but also very interesting possibility~\cite{Sakurai:2023nsc,Morales:2024jhj}.
However, more work on the theory side is required in order to better understand and measure relativistic multi-particle quantum correlations.

\end{itemize}

\FloatBarrier

\section{Conclusion}
\label{sec:conc}

The observation of entanglement and studies of other quantum correlations in collider experiments mark a significant step toward integrating QIT with HEP.  While initial studies at the LHC have demonstrated the feasibility of such measurements, they have also highlighted key challenges, including the need for improved event reconstruction, more precise Monte Carlo simulations, a reduction of systematic uncertainties and larger datasets. Addressing these issues will be crucial for refining our ability to extract quantum observables and for expanding the range of processes where quantum effects can be explored.

Future collider experiments will offer new opportunities to probe entanglement, Bell nonlocality, and other quantum phenomena with greater precision. 
The increased luminosity at the HL-LHC is expected to enable the first measurement of Bell nonlocality in top-pair production and potentially also in Higgs decays. 
Meanwhile, lepton colliders would allow for unprecedented tests of quantum correlations in electroweak processes, such as $e^+e^- \to \tau^+\tau^-$ and Higgs production, benefiting from precise control over initial-state conditions and reduced backgrounds.  These facilities will also offer a unique opportunity to explore new physics beyond the Standard Model, as quantum information techniques could help identify subtle signals of new particles or interactions.

Looking further ahead, reaching center-of-mass energies of dozens of TeV in future hadron colliders will provide a unique environment to study multipartite entanglement in high-energy collisions, where the production of multiple heavy particles will become more frequent. 
This will allow for systematic studies of entanglement in complex final states, which could reveal new patterns of quantum correlations. 
Additionally, future colliders will enhance our sensitivity to new physics by probing quantum observables in regions of phase space previously inaccessible. By pushing the boundaries of both experimental techniques and theoretical models, future studies will deepen our understanding of QIT in HEP and provide new avenues for exploring the Standard Model and beyond.

\clearpage

\section*{Acknowledgments}
YA is supported by the National Science Foundation under Grant No.\ PHY-2310094. FF acknowledges financial support by the MSCA European Fellowship QUANTUMLHC program Horizon Europe G.A.101107121 CUP J33C23001080006. ML is supported by U.S. NSF Grant No. \ PHY-2412696 and by U.S. DOE Grant No. \ DE-SC0007914. LM is supported by the Estonian Research Council under the RVTT3, TK202 and PRG1884 grants. The work of JAAS, AB, JAC and JM has been supported by the Spanish Research Agency through projects PID2022-142545NB-C21,  PID2022-142545NB-C22 and CEX2020-001007-S funded by MCIN/ AEI/ 10.13039 / 501100011033. The work of AB is in addition supported through the FPI grant PRE2020-095867 funded by MCIN/AEI/10.13039/501100011033. The work of AJB is funded in part via Science and Technology Facilities Council (STFC) grants ST/R002444/1 and ST/S000933/1, and the John Templeton Foundation Grant No. 63206. BF has been supported by Grant ANR-21-CE31-0013 from the French \emph{Agence Nationale de la Recherche}. DG acknowledges support by US Department of Energy Grant Number DE-SC 0016013. The work of TJH at Argonne National Laboratory was supported by the U.S. Department of Energy under contract DE-AC02-06CH11357. PH acknowledges support by the National Science Centre in Poland through the research grant Maestro (2021/42/A/ST2/0035). JH is supported by The Royal Society URF\textbackslash R241013, URF\textbackslash R1\textbackslash191524. FM and MS~acknowledge financial support through the the project QIHEP - financed through the PNRR, funded by the European Union – NextGenerationEU, in the context of the extended partnership PE00000023 NQSTI - CUP J33C24001210007. FM and PL are also supported by PRIN2022 Grant 2022RXEZCJ, funded by the European Union – NextGenerationEU. The work of JM was supported by the Royal Society grant URF\textbackslash R1\textbackslash201519 and STFC grant ST/W000512/1. RAM acknowledges financial support by CONICET and FONCyT through the project PICT-2021-00374. JRMdN is supported by Spain’s Agencia Estatal de Investigacion Grant No. PID2022-139288NBI00. DP~acknowledges financial support by the Italian Ministry of University and Research (MUR) through the PRIN2022 Grant 2022EZ3S3F. Several authors acknowledge financial support from the COMETA COST Action CA22130, funded by COST (European Cooperation in Science and Technology). G.P.~acknowledges financial support from the MUR under the PRIN2022 funding scheme Grant No.~20229KEFAM, funded by the European Union – NextGenerationEU.  ST was supported by the Office of High Energy Physics of the U.S. Department of Energy (DOE) under Grant No.~DE-SC0012567, and by the DOE QuantISED program through the theory consortium “Intersections of QIS and Theoretical Particle Physics” at Fermilab (FNAL 20-17). MV~acknowledges financial support by the Spanish ministry of science under grant PID2021-122134NB-C21 and by the government of the Valencia region under grant CIPROM/2021/073. CDW~is supported by the UK STFC Consolidated Grant ST/P000754/1 "String theory, gauge theory and duality". MJW is supported by the Australian Research Council grants CE200100008 and DP220100007. KZ~is supported by the U.S. Department of Energy under Grant No.~DE-SC0007881.

\clearpage

\bibliographystyle{JHEP}
\bibliography{main.bib}

\end{document}